# Adhesion and non-linear rheology of adhesives with supramolecular crosslinking points†

X. Callies,*[a] C. Fonteneau,[b] S. Pensec,[b] L. Bouteiller,[b] G. Ducouret[a] and C. Creton*[a]

Soft supramolecular materials are promising for the design of innovative and highly tunable adhesives. These materials are composed of polymer chains functionalized by strongly interacting moieties, sometimes called "stickers". In order to systematically investigate the effect of the presence of associative groups on the debonding properties of a supramolecular adhesive, a series of supramolecular model systems has been characterized by probe-tack tests. These model materials, composed of linear and low dispersity poly(butylacrylate) chains functionalized in the middle by a single tri-urea sticker, are able to self-associate by six hydrogen bonds and range in molecular weight ($M_n$) between 5 and 85 kg mol$^{-1}$. The linear rheology and the nanostructure of the same materials (called "PnBA3U") were the object of a previous study. At room temperature, the association of polymers *via* hydrogen bonds induces the formation of rod-like aggregates structured into bundles for $M_n$ < 40 kg mol$^{-1}$ and the behavior of a soft elastic material was observed ($G' \ll G''$ and $G' \sim \omega^0$). For higher $M_n$ materials, the filaments were randomly oriented and the polymers displayed a crossover towards viscous behavior although terminal relaxation was not reached in the experimental frequency window. All these materials show, however, similar adhesive properties characterized by a cohesive mode of failure and low debonding energies ($W_{adh}$ < 40 J m$^{-2}$ for a debonding speed of 100 μm s$^{-1}$). The debonding mechanisms observed during the adhesion tests have been investigated in detail with an Image tools analysis developed by our group. The measure of the projected area covered by cavities growing in the adhesive layer during debonding can be used to estimate the true stress in the walls of the cavities and thus to characterize the *in situ* large strain deformation of the thin layer during the adhesion test itself. This analysis revealed in particular that the PnBA3U materials with $M_n$ < 40 kg mol$^{-1}$ soften very markedly at large deformation like yield stress fluids, explaining the low adhesion energies measured for these viscoelastic gels.

## Introduction

The interest in developing innovative and smart materials from supramolecular polymers[4,5] has considerably grown in the last few years. These materials are composed of polymer chains functionalized by strongly interacting moieties, often called "stickers". Among the numerous studies focusing on the properties of these new materials,[6] some of them reported promising results in the field of adhesion, especially with the emergence of non-sticky and self-healing materials[7–9] or stimuli-responsive adhesives.[10] These studies raised fundamental questions about the effect of stickers on the adhesive properties of supramolecular materials, *via* their influence on the surface energy[11] or the rheology of these materials.

Since the first investigation on ionomers in the 1960s, it is well known that the non-covalent association of stickers in the melt can lead to a temporary network *via* the formation of dimers, clusters or aggregates.[12,13] These physical cross-links strongly increase the relaxation time of the polymer chains and considerably affect their rheological properties in the linear and non-linear regimes.[14] That is why weakly interacting moieties, such as acrylic acid, are commonly used to adjust the properties of Pressure Sensitive Adhesives (PSAs[15]), the rheological properties of which are essential for their applications.[16] These soft adhesives are thin films that stick on almost any surface by simple contact and that can be ideally detached from the substrate without any residues. In order to have an optimized combination of resistance to shear and "tacky" character, PSAs[15] require very well defined viscoelastic properties in order to reach large bulk deformations and thus, high debonding energies. Although functional groups are usually used in the elaboration of PSAs, few systematic studies were carried out on well-defined supramolecular model systems.[17,18]

The general strategy of our project is to adjust the adhesive properties of unentangled or middly entangled mono- or multi-functional polymers by varying the strength and the density of stickers. In a previous study on short poly(isobutene) chains center-functionalized by a single hydrogen bonding sticker, some of us showed promising viscoelastic properties to obtain adhesives.[17] Although a gel-like behavior was observed at small deformation, the dynamic character of the supramolecular interactions could lead to highly dissipative properties when deformed at large strain. In order to investigate more systematically the effect of the bulk linear rheology on the adhesion of such supramolecular materials, the present paper focuses on the adhesive strength of model supramolecular materials, previously well-characterized in structure and linear rheology.[1,2] The model materials, called "PnBA3U", are composed of poly(butylacrylate) chains with a narrow dispersity (see Table 1) and functionalized in their center by a single tri-urea sticker able to interact by six hydrogen bonds[19] (see Fig. 1). The molecular weight of the polymers ($M_n$) was systematically varied from 5 to 85 kg mol$^{-1}$, decreasing the density of the stickers $\Phi_s$ from 7 to 0.4%. Below a critical molecular weight $M_c \sim 40$ kg mol$^{-1}$, the self-assembly of stickers into rod-like aggregates structured into bundles[2] prevents the polymer chains from flowing at room temperature and a soft elastic response is observed in the linear regime[1] (see Fig. 9 in the discussion part). At higher molecular weights, stickers are too dilute to form locally oriented rods and PnBA3Us behave like viscoelastic fluids.

[a] Sciences et Ingénierie de la Matière Molle, CNRS UMR 7615, ESPCI Paris, PSL Research University, 10 rue Vauquelin, F-75231 Paris cedex 05, France. E-mail: xavier.callies@etu.upmc.fr, costantino.creton@espci.fr

[b] Sorbonne Universités, UPMC Univ Paris 06, CNRS, IPCM, Chimie des Polymères, F-75005 Paris, France

† Electronic supplementary information (ESI) available. See DOI: 10.1039/c6sm01154c

Measurements of linear viscoelasticity are rapid and well-established for the characterization of the small strain properties of materials and so, they are commonly used to study the effect of the dynamics of stickers in the literature. This article focuses, however, on the essential role played by the non-linear rheology in controlling the adhesive properties of the materials as tested in classical debonding tests such as probe-tack tests.[20] The adhesion tests on PnBA3U were carried out on a home-made set-up[21] fitted with a microscope which allowed us to characterize the *in situ* deformation of the thin layer during the probe-tack test itself. The debonding mechanisms at the interface between the soft adhesive layer and the rigid substrate was directly observed and recorded. From the analysis of the images,[3] the true extensional stress is calculated and the behavior of the fibrils between the cavities formed upon debonding could thus be characterized in an analogous way as in classic tensile tests, albeit with a variable applied strain rate. This information is important to understand the complex mechanisms behind the probe-tack curves.

In this article, the experimental methods are first presented and the stress–strain curves of the probe-tack tests are then described in the light of the images of the debonding mechanisms. In the Analysis and discussion part, these pictures are exploited to highlight the effect of non-linear rheology on the adhesive properties.

Table 1 Chemical parameters of PnBA3U systems

| Samples | $M_n$ (g mol$^{-1}$) | $Đ$ | DP | $\Phi_s$ (%) |
|---|---|---|---|---|
| PnBA3U5 | 5200 | 1.24 | 36 | 6.8 |
| PnBA3U8 | 8000 | 1.2 | 58 | 4.4 |
| PnBA3U12 | 12 000 | 1.24 | 89 | 3.0 |
| PnBA3U18 | 18 000 | 1.23 | 140 | 1.9 |
| PnBA3U40 | 40 000 | 1.32 | 310 | 0.89 |
| PnBA3U85 | 85 000 | 1.34 | 660 | 0.42 |

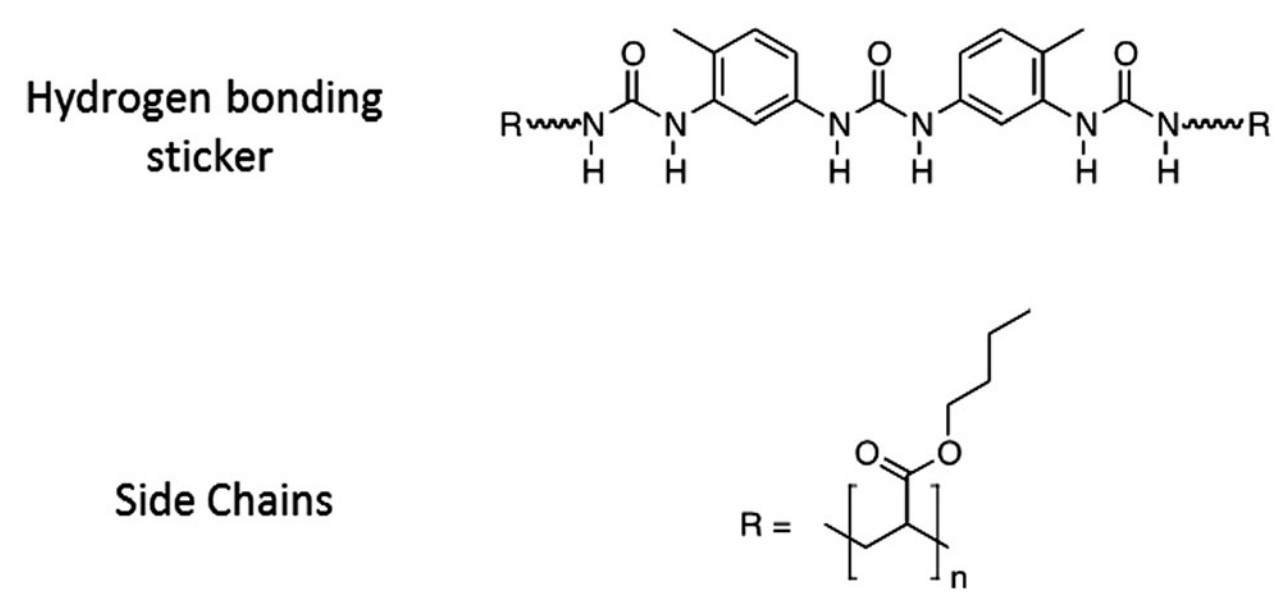

Fig. 1 Schematic chemical structure of the PnBA3U supramolecular polymer, the polar sticker is linked to two linear poly-*n*-butyl acrylates (R). The degree of polymerization changes from one material to another.

## Experimental part

### Materials

A series of poly-*n*-butyl acrylates of increasing and well-defined molecular weight, center-functionalized with a tri-urea sticker, was synthesized as previously described by Fonteneau and coworkers[22] (see Fig. 1). The molecular characteristics of the supramolecular polymers investigated in this paper are reported in Table 1. The number average molar mass $M_n$ and molar mass distribution $Đ$ ($Đ = M_w/M_n$ with the weight average molar mass $M_w$) were determined by size exclusion chromatography (SEC) in tetrahydrofuran, using a refractive index detector and a polystyrene calibration curve for samples PnBA3U5, PnBA3U12 and PnBA3U18 or with a triple detection set-up for the other samples (see ESI of ref. 1 for representative SEC curves). The number average degree of polymerization DP and the sticker density $\Phi_s$ in the polymer matrix are then estimated: DP = $(M_n - M_s)/M_{bu}$ and $\Phi_s = M_s/M_n$, by using the molar mass $M_s$ of the sticker and the molar mass $M_{bu}$ of the butylacrylate monomers. In order to check the absence of any residual solvent in PnBA3U after synthesis, samples are analyzed by $^1$H NMR (Bruker Avance 200).[1]

### Adhesion tests

The probe-tack test[20] as we carried it out, consists of bringing the surface of a solid probe into contact with the thin adhesive layer coated on a rigid substrate and in measuring the force $F_T$ required to detach it at a constant debonding speed $V_{deb}$ (see Fig. 2). Adhesion tests were carried out on thin (100 μm) adhesive layers deposited on glass slides (2.6 × 10 × 0.2 cm$^3$, purchased from Preciver). Before use, glass slides were cleaned with ethanol and acetone. In order to get a reproducible film of thickness 100 μm ± 10 μm, each supramolecular polymer was deposited by slow evaporation of a solution in toluene under a glass cover during two days followed by a final heat treatment at 70 °C in an oven under reduced pressure (~200 mbar) for two additional days. The solution was prepared by dissolving

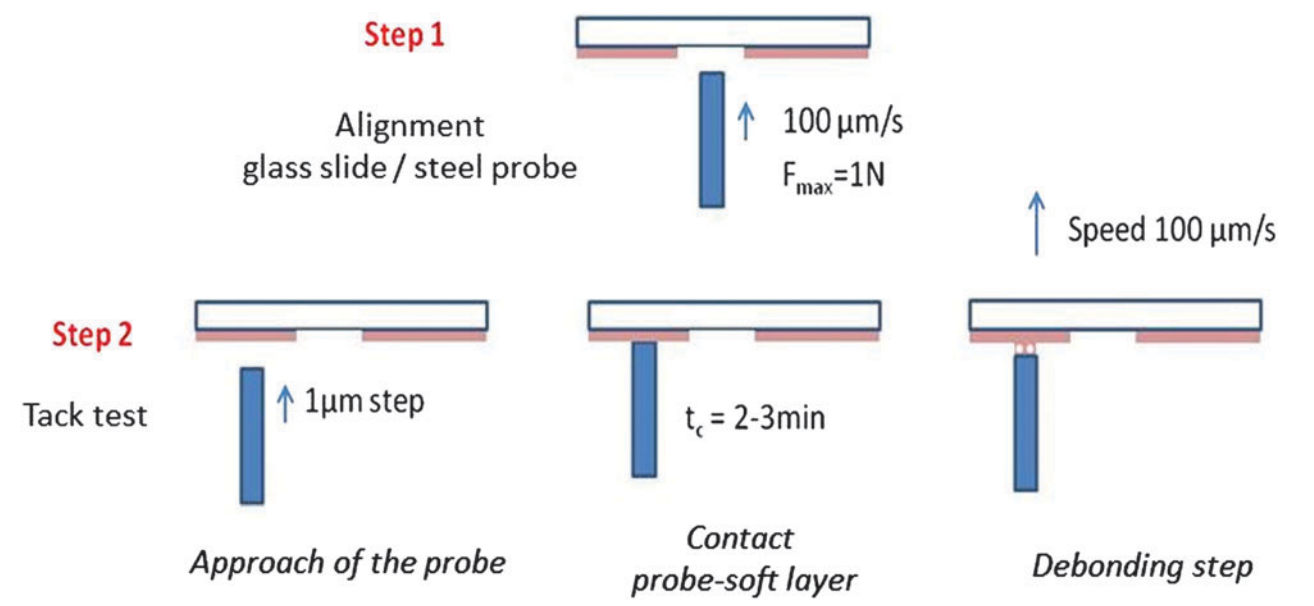

Fig. 2 Experimental procedure for the adjustment of the glass slide/probe alignment (step 1) and the tack test (step 2).

300 mg of supramolecular material in 2 mL of toluene. The dry film thickness was measured by a white light scanning technique using an optical profilometer (Microsurf 3D, Fogale nanotech). Due to the limited control of the evaporation process and, thus, of the thickness over the same glass slide, several measurements were necessary for each sample. In order to get a perfectly smooth and reflective surface, the stainless steel probes (diameter $\Phi$ = 5.95 ± 0.02 mm) were mechanically polished. After each probe-tack test, the probe is cleaned with acetone.

Adhesive tests were carried out at room temperature with a home-made probe-tack set-up described in detail elsewhere.[21] Unlike commercial set-ups for probe tack tests, the debonding of the probe from the adhesive layer is observed through an optical microscope with a long-working distance lens (1.25× to 20×) and the displacement of the probe is controlled using micrometric stepping motors[21] (see Fig. S1, ESI†). Micrometric stepping motors are used to control the vertical motion of the probe and thus the thickness of the adhesive layer during the approach and the debonding step, and three screws are used to adjust the coplanarity of the probe and the film. Precise control of the geometric parameters is required to obtain reproducible results for liquids.

The displacement $d(t)$ of the probe relative to the upper-plate is measured using a fiberoptic sensor (Philtec D63 LPT) and a mirror screwed on the probe with a precision of 0.2 µm. The load cell (40 N ± 0.1 N) required for the measurement of the normal force $F_T$ is positioned in series with the probe. The compliance $K$ = 5.5 µm N$^{-1}$ of the apparatus measured by a blank test (without any viscoelastic layers on the glass slide) is used to calculate the thickness of the adhesive layer $h(t)$ at an instant $t$ during the adhesion test:

$$h(t) = d(t) - K \times F_T(t) + h_0 \qquad (1)$$

with $h_0$ being the initial thickness of the adhesive film. The stepping motors (PI Instruments Karlsruhe, Germany) and the separate acquisition card (Data Translation DT301, Malboro, USA) are controlled using a National Instruments Labview software module developed in-house. The experimental data are the force, the displacement of the motors and the displacement measured by the optical sensor as a function of time. The data file is synchronized with the video capture of the test using the CCD camera (connected to the optical microscope) *via* an outside timer.

The alignment between the probe and the glass slide is achieved by optimizing the contact surface between the probe and a commercial acrylic PSA through the microscope. Before each probe-tack experiment on a thin viscoelastic layer, the coplanarity of the glass slide and the steel probe is checked in the middle of the glass slide where the adhesive layer has been removed (see step 1 in Fig. 2). A good alignment is confirmed by the observation of optical fringes. Fine control of the probe motion is required to know precisely the thickness of the adhesive layer at the beginning of the debonding step and thus a good reproducibility of the results (see Fig. S2, ESI†). Unlike the classical probe-tack test where the probe stops at a fixed value of the compression force, the probe is approached by 1 µm steps under the control of the experimenter (see step 2 in Fig. 2). The probe is stopped when the contact surface is large enough (>60% of the surface of the probe). This procedure is necessary for having good control of the initial thickness of the sample. Manual control of the probe translation by the experimenter and the stabilization of the contact surface prevents the precise control of the contact time $t_c \sim 3$ min and also prevents the use of much shorter contact times (<30 s) with the present set-up. After waiting approximately $t_c \sim 3$ min, the glass slide is driven up at a controlled debonding speed (see the picture in the ESI†). No significant effect was observed when $t_c$ was increased up to $t_c \sim 15$ min. The adhesion tests were carried out at room temperature, *i.e.* between 23 °C and 25 °C. The controlled debonding velocity can be fixed from 1 µm s$^{-1}$ to 100 µm s$^{-1}$ but only adhesion tests at 100 µm s$^{-1}$ are reported in this article.

Probe-tack curves conventionally plot the nominal stress $\sigma_0(t) = F_T(t)/S_0$ *versus* the nominal strain $\varepsilon_0(t) = (h(t) - h_0)/h_0$ or the stretching ratio $\lambda_0(t) = 1 + \varepsilon_0(t)$, where $S_0$ is the contact area between the probe and the adhesive layer at the maximum compression stage. The area under the $\sigma_0 = f(\varepsilon_0)$ curve is used to calculate the debonding energy of the soft layer, $W_{adh}$ (J m$^{-2}$), *i.e.* the energy necessary to detach the probe from the layer for the given traction speed:

$$W_{adh} = h_0 \int \sigma_0 d\varepsilon_0 \qquad (2)$$

### Methods for image analysis

Tanguy and coworkers[3] recently proposed to characterize the *in situ* deformation of the thin layer during the probe-tack test from the analysis of the debonding pictures (recorded using a CCD camera). As explained in detail by these authors,[3] the first step of this method consists of converting the debonding images into binary pictures, where bubbles and fibrils can be clearly identified (see Fig. 3A and C). The second step consists of detecting each bubble and in measuring its projected area, as illustrated in Fig. 3B. The total projected area covered by the cavities, *i.e.* the sum of the projected area of all bubbles, is denoted as $A_b$. As each cavity is defined by a convex envelope, $A_b$ drops to zero at the equilibration of pressure, *i.e.* when all bubbles are coalesced. The third step consists of measuring the projected area of the fibrils, denoted as $A_e$ (the white area in Fig. 3C). The measurement of $A_e$ from the images can be used to calculate an effective elongation $\langle\lambda\rangle$ along the tensile direction in the thinnest cross-section of the walls by assuming a deformation of the wall at constant volume.[3]

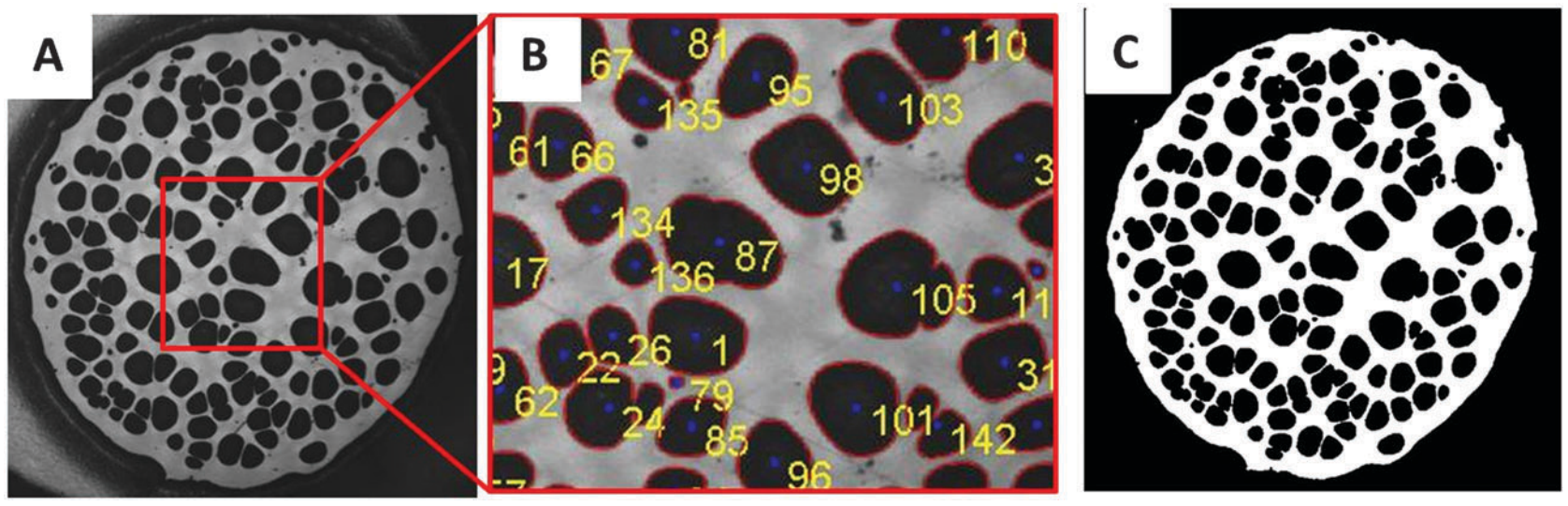

Fig. 3 (A) Picture of the debonding patterns observed for PnBA3U5 at a traction speed $V_{deb}$ = 100 μm s$^{-1}$. (B) Focus on the previous image and identification of the cavities by Matlab© image analysis. (C) Conversion of the image (A) into a binary picture.

$$\langle\lambda\rangle = \frac{A_0}{A_e} \tag{3}$$

With the same information, the true strain $\varepsilon_H$ can be calculated from $A_e$:

$$\delta\varepsilon_H = -\frac{\delta A_e}{A_e} \tag{4}$$

$$\varepsilon_H = \ln\langle\lambda\rangle \tag{5}$$

As illustrated by the scheme in Fig. 4A, the deformation of the thin film during debonding induces the formation of a disc with a fibrillar structure. As this disc is being stretched in the tensile direction, its section $A_t$ decreases and can be monitored by considering the sum of $A_e$ and $A_b$, $A_t = A_e + A_b$.

In addition to the effective elongation, the analysis of the debonding video synchronized with the force sensor can be used to calculate true stress $\sigma_e$, *i.e.* the normal component of the stress in the walls between cavities before the equilibration of pressure as well as in the remaining fibrils after the equilibration of pressure.

$$\sigma_e = \frac{F_m}{A_e} \tag{6}$$

In the first case, the respective force $F_m$ is obtained by subtracting the force $F_p$ due to work against the atmospheric pressure from the measured force $F_T$. The force $F_p$ is directly proportional to the difference between the external pressure $P_{atm}$ and the pressure $P_b$ inside the bubbles. After the nucleation step, the growth of a cavity highly modifies the distribution of stress around the cavity, limiting the growth of the neighboring cavities.[23,24] The total area covered by the cavities, $A_b$, but also the projected area of fibrils between cavities must thus be taken into account to estimate $F_p$. This can be done by considering the convex envelope[3] $A_c$ around cavities, as shown in Fig. 4.

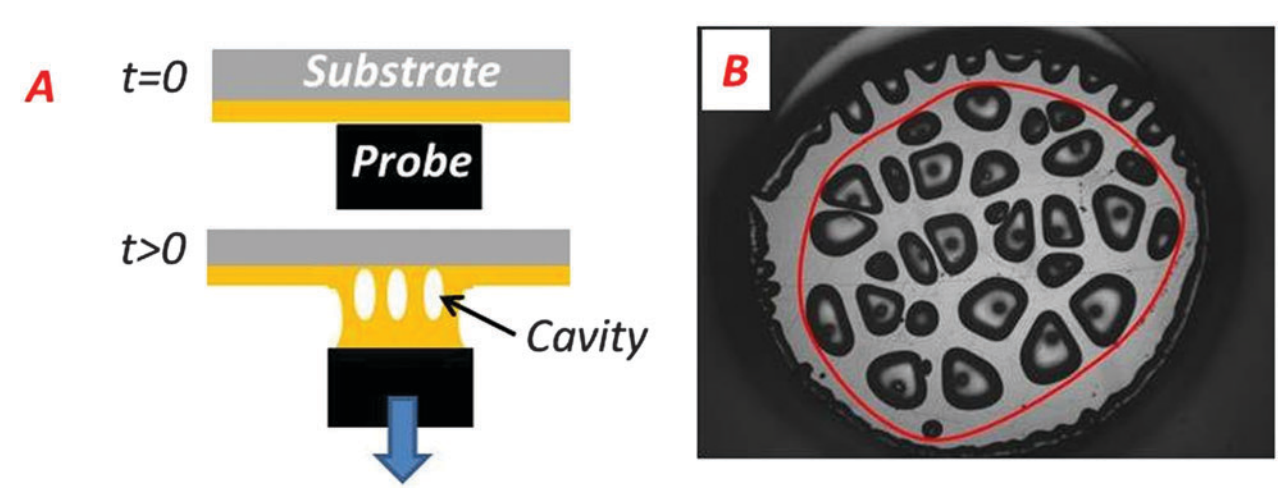

Fig. 4 (A) Scheme of the side-viewed fibrillar structure formed by the walls between cavities. (B) Convex envelope of the region occupied by cavities (red solid line).

$$F_p = A_c(P_{atm} - P_b) \tag{7}$$

As the order of magnitude of $P_b$ is that of the vapor pressure,[25] $P_{atm} \gg P_b$ and thus, $F_p$ can be simply approximated:

$$F_p \approx A_c P_{atm} \tag{8}$$

The effective stress can then be estimated:

$$\sigma_e = \frac{F_T - A_c P_{atm}}{A_e} \tag{9}$$

In the case of viscoelastic fluids, the elongational flow in the fibrils can be characterized by an approximate local transient elongational viscosity $\eta^+$ *via* the Hencky strain rate $\dot{\varepsilon}_H$:

$$\eta^+ = \frac{\sigma_e}{\dot{\varepsilon}_H} \tag{10}$$

The delimitation of the convex area $A_c$ requires a good visual contrast between cavities and fibrils in order to detect precisely the edges of cavities. $A_c$ is particularly difficult to measure precisely during the phase of equilibration of pressure, when the walls between cavities are very thin (*e.g.* in the highlighted $\lambda$ region of the graph $\sigma_{0f} = (\lambda_0)$ in Fig. 6). The effective stress and the elongational viscosity measured during this phase, *via* eqn (9) and (10), are quite noisy and will be not discussed in the present work.

After the coalescence of all bubbles, $F_p = 0$ and the effective stress can be simply calculated using the projected area of fibrils $A_e$ and the force $F_T$:

$$\sigma_e = \frac{F_T}{A_e} \tag{11}$$

## Results

Representative stress–strain curves of probe tack tests obtained for 100 μm thick PnBA3U films with a debonding velocity of 100 μm s$^{-1}$ are shown in Fig. 5. For all PnBA3Us, stress–strain curves show a stress peak at small strain, followed by a continuous decrease until a sharp decrease from $\sigma_0 \approx 10^5$ Pa to $\sigma_0 \approx 10^4$ Pa. At larger deformations, $\sigma_0$ continues to decrease more or less steeply until the complete detachment of the probe. The stress variation in this final step is clearly different for the different adhesive layers and is more progressive for the highest molecular weight materials, as shown by the lines drawn in Fig. 5.

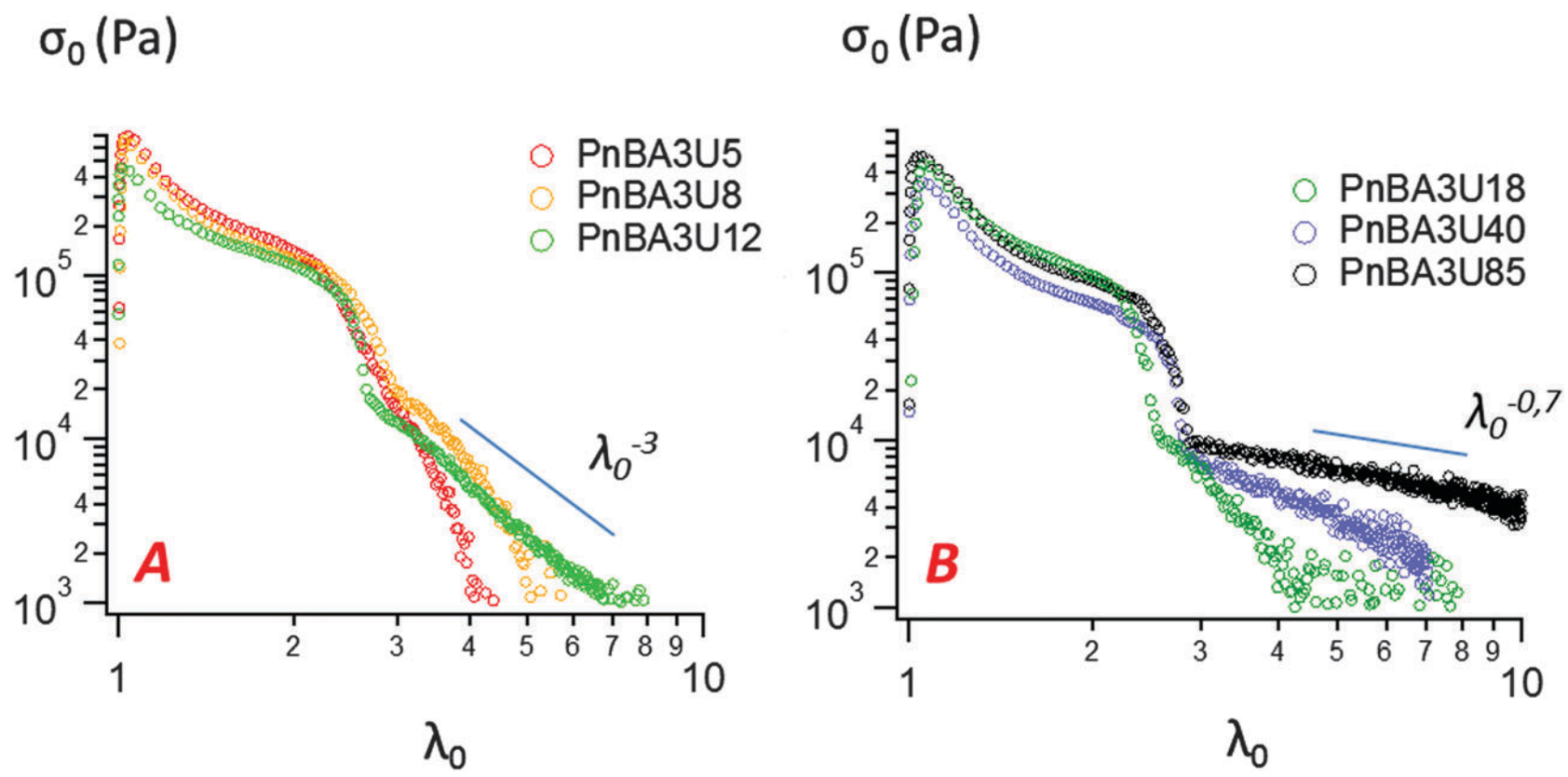

Fig. 5 Probe-tack curves for low (A) and high (B) molecular weight PnBA3U polymers at room temperature (traction speed 100 μm $s^{-1}$).

In order to interpret this change in the shape of the stress–strain curves with $M_n$, the experimental force and displacement data were correlated with the images of the debonding process taken using a camera synchronized with the stepping motors. Although debonding patterns may appear qualitatively similar to the untrained eye, two types of debonding processes could be identified over the investigated $M_n$ range and are illustrated using PnBA3U8 and PnBA3U40 in Fig. 6.

The first type of debonding mechanism occurring in our study is observed in the $M_n$ range from 5 up to 18 kg $mol^{-1}$ and is illustrated using PnBA3U8 in Fig. 6. For PnBA3U8, the comparison of Fig. 6A and B reveals the rapid macroscopic growth of micrometric bubbles at small strain ($\lambda_0 < 1.2$). This cavitation process corresponds to the high stress peak observed on the stress–strain curve. The decrease of the nominal stress that follows results from the lateral growth of the bubbles and from the decrease of the thickness of the walls between them (see Fig. 6C). The external walls at the periphery of the contact zone between the adhesive and the probe do not show any Saffman–Taylor instabilities and break at $\lambda_0 \approx 2.2$ due to the difference in hydrostatic pressure in the ambient atmosphere[3,25] (see Fig. 6D). The breakage of the walls between cavities from the periphery occurs when the cavities expanding toward the external wall touch and coalesce with the wall. The atmospheric pressure then propagates toward the center of the probe in the range of $\lambda_0$ which is highlighted on the graph $\sigma_0 = f(\lambda_0)$ in Fig. 6. After the coalescence of the bubbles, the decrease in stress results from the elongation of the residual fibrils between the probe and the glass slide (Fig. 6E). At the end of the adhesion test, fibrils break and residues remain on the steel probe (see Fig. 7). The mode of failure is called "cohesive".

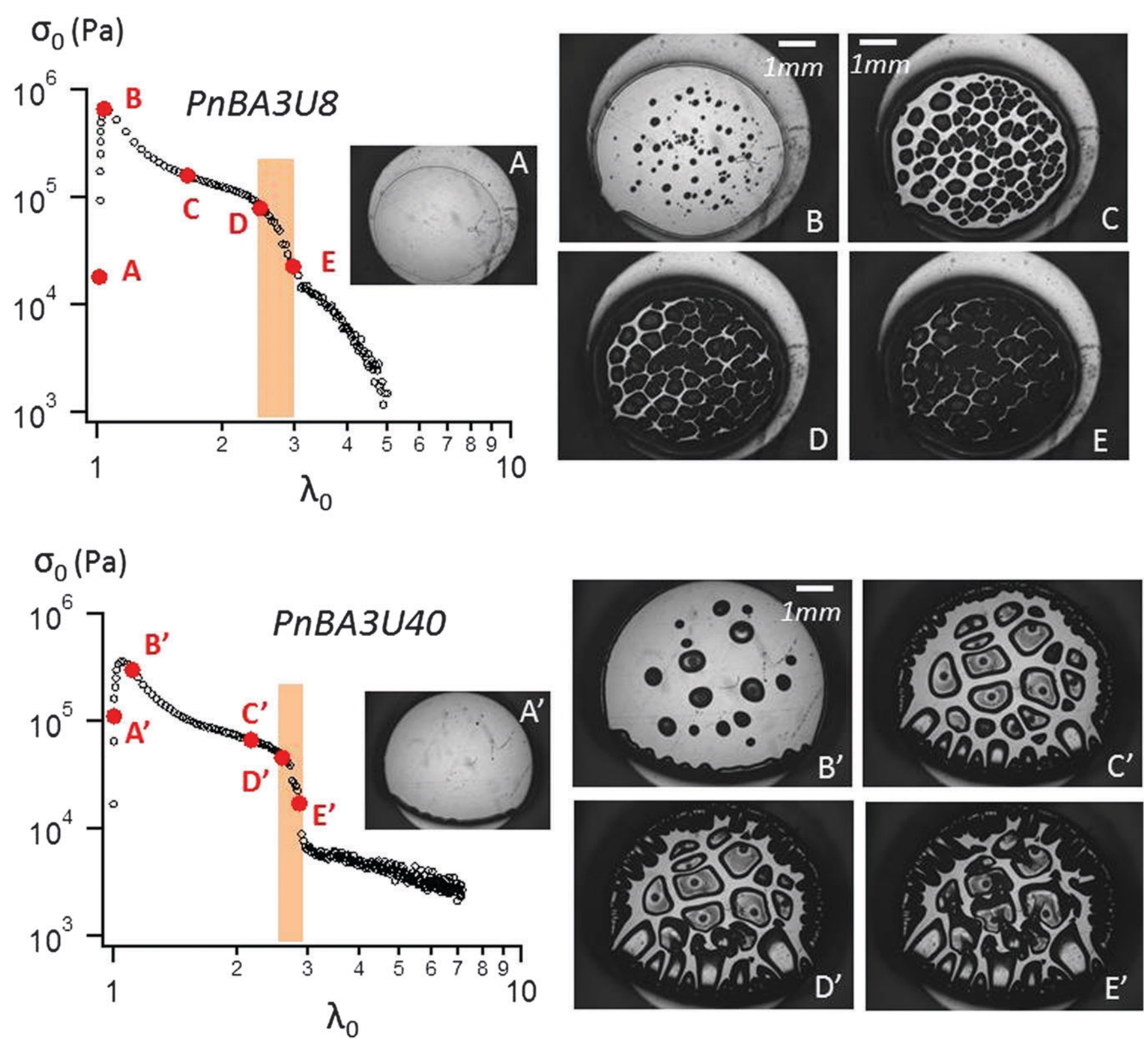

Fig. 6 Probe-tack curve for PnBA3U8 (top) and PnBA3U40 (bottom) with correlated pictures at 100 μm $s^{-1}$.

The second type of debonding process is observed for $M_n \geq 40$ kg mol$^{-1}$ and is illustrated with PnBA3U40 in Fig. 6. For PnBA3U40, the stress peak results from the growth of micro-bubbles as well as the growth of Saffman–Taylor instabilities from the edge of the contact zone (see Fig. 6A′ and B′) as commonly observed for viscoelastic fluids.[26] As the nominal stress decreases continuously, fingers and bubbles are observed to grow (see Fig. 6C′ and D′) up to their intersection which induces the equilibration of the pressure and the nominal stress drops at $\lambda_0 \approx 2.9$. After the coalescence of the cavities, residual fibrils are stretched (see Fig. 6E′) and finally break at much larger values of $\lambda_0$. Like for low molecular weight materials, the eventual mode of failure is clearly cohesive (see Fig. 7).

The area under the stress–strain curve can be used to measure the energy $W_{adh}$ per unit of contact area required to detach the stainless steel probe from the thin adhesive layer. As shown in Fig. 8, the variation of $W_{adh}$ and that of the maximal stress $\sigma_{0,max}$ with molecular weight are similar. When $M_n$ increases from 5 to 40 kg mol$^{-1}$, $W_{adh}$ and $\sigma_{0,max}$ continuously decrease, from 36 ($\pm$2) J m$^{-2}$ to 22 ($\pm$2) J m$^{-2}$ and from 700 ($\pm$100) kPa to 350 ($\pm$50) kPa respectively. Both parameters increase for $M_n$ = 85 kg mol$^{-1}$, $W_{adh}$ = 28 ($\pm$2) J m$^{-2}$ and $\sigma_{0,max}$ = 470 ($\pm$65) kPa.

In order to characterize the elongation of the film, two other parameters are also plotted in Fig. 8: the stretching ratio $\lambda_{0,p}$ at the equilibration of the pressure (*i.e.* $\sigma_0 \approx 10$ kPa) and the stretching ratio $\lambda_{0,f}$ at the minimum of stress detected by the load cell (*i.e.* $\sigma_0 \approx 1$ kPa). In the investigated molecular weight range, $\lambda_{0,p}$ varies in a narrow window, with a minimum $\lambda_{0,p}$ = 2.4 ($\pm$ 0.2) measured at 18 kg mol$^{-1}$ and a maximum $\lambda_{0,p}$ = 3.1 ($\pm$0.2) measured at 8 kg mol$^{-1}$. However, the molecular weight dependence is much more pronounced for $\lambda_{0,f}$. While $\lambda_{0,f}$ remains quasi-constant and low (<5) for the supramolecular gels, $\lambda_{0,f}$ highly increases with the length of the polymer chains for the supramolecular viscoelastic liquids ($M_n \geq 40$ kg mol$^{-1}$) demonstrating the much more stabilizing effect on the stretched fibrils of the polymer entanglements relative to the supramolecular associations between short chains.

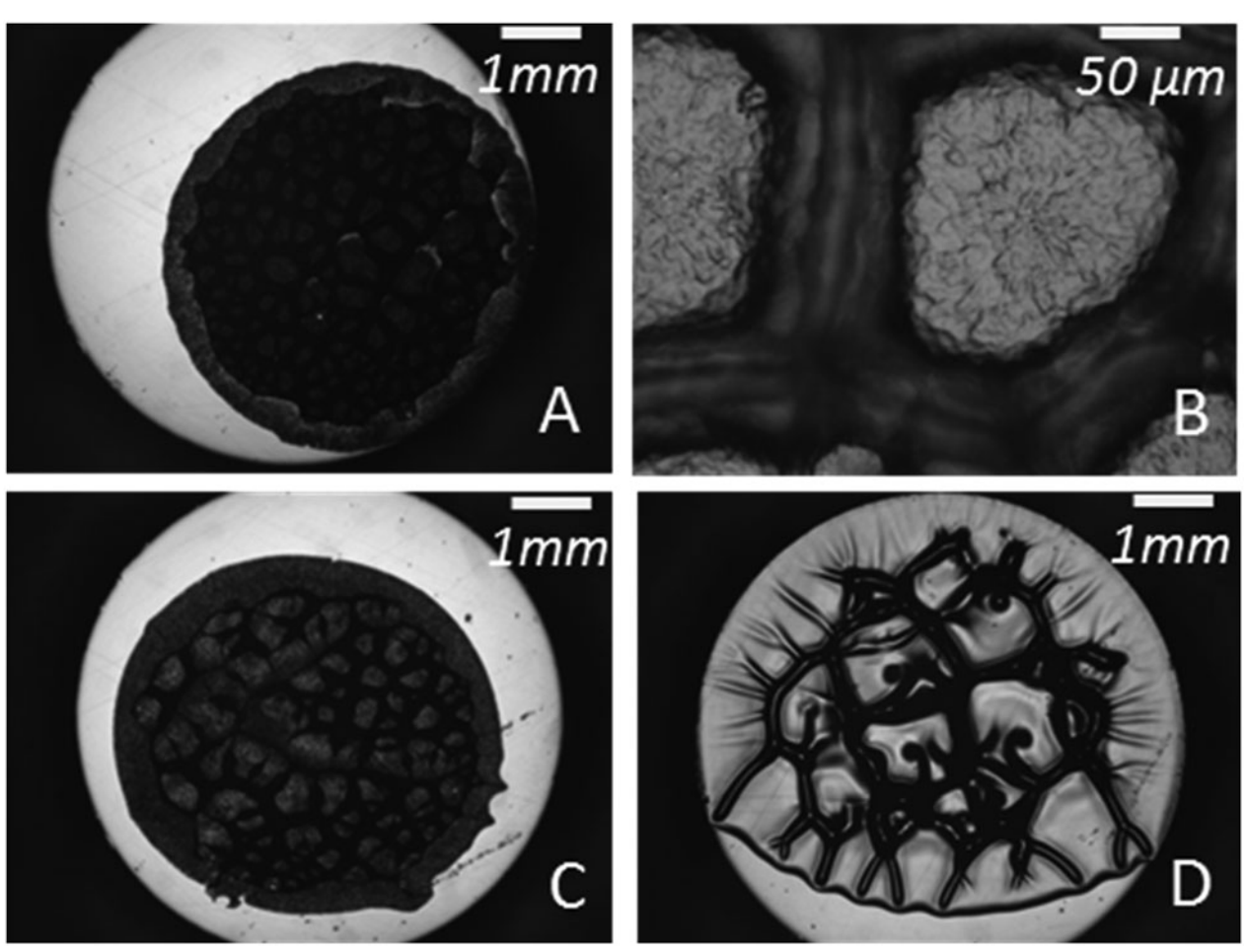

Fig. 7 Images of steel probes after tests for PnBA3U5 (A and B), PnBA3U8 (C) and PnBA3U40 (D).

Some tests were carried out at lower debonding speeds (at 10 and 1 µm s$^{-1}$) and they qualitatively showed a similar weak molecular weight dependence of the debonding energy and a cohesive mode of failure, albeit values of adhesion energy were lower. These experiments were carried out using a less accurate protocol and are not reported here.

## Analysis and discussion

### Linear rheology and probe-tack tests

The most striking aspect of our experimental study is the liquid-like behavior of all PnBA3U materials in the probe-tack tests, while a solid-like response was clearly observed for low molecular weight materials ($M_n \leq 20$ kg mol$^{-1}$) in linear rheology (see Fig. 9). For all investigated molecular weights, we observed characteristic features of viscoelastic fluids[27] in the probe tack tests, such as the cohesive mode of fracture as well as the cavitation and fibrillation processes as illustrated in Fig. 6 and 7. The typical shape of the stress–strain curves, characterized by a stress peak followed by a continuous decrease in nominal stress, is due to the low compliance of the confined adhesive layer at the beginning of the probe-tack test.[25] When the cavities appear and the thickness of the layer increases, its compliance increases and becomes higher than that of the apparatus. As a result the spring force induced by the deformed apparatus[28,29] pulls on the layer and the stress decreases strongly, as observed in Fig. 5 and 6. At this stage, the cavities rapidly grow and the walls between them are stretched before finally breaking. The sharp contrast of rheological behaviors observed at small and large deformation is best revealed by comparing the results with different predictions from existing adhesion theories based on linear rheology.

A first difference between PnBA3U materials and more classic soft adhesives is the condition for nucleation of cavities characterized by the high stress peak. In elastic materials such as cross-linked rubbers,[30,31] the growth of an existing cavity (typically at the surface of the probe) occurs when the hydrostatic pressure in the thin layer greatly exceeds its elastic modulus measured in the linear regime. As the critical stress for the growth of cavities can be difficult to estimate for viscoelastic materials, Lakrout *et al.*[27] proposed to consider the peak nominal stress $\sigma_{0,max}$. In the case of acrylic Pressure Sensitive Adhesives, these authors observed that the stress peak was proportional to the linear storage modulus $G'(\omega_c)$: $\sigma_{0,max} \approx 10G'(\omega_c)$ measured at the representative angular frequency $\omega_c = 2\pi V_{deb}/h_0$ of the adhesion test. For PnBA3Us, $\sigma_{0,max}$ is also observed to be proportional to $G'(\omega_c)$ for $M_n \geq 12$ kg $mol^{-1}$: $\sigma_{0,max} \approx 20G'(\omega_c)$ (see Fig. 10). However, this linear relationship clearly fails for PnBA3U8 and PnBA3U5, where the stress peak observed in probe tack tests is much lower than the peak predicted from linear rheology.[27]

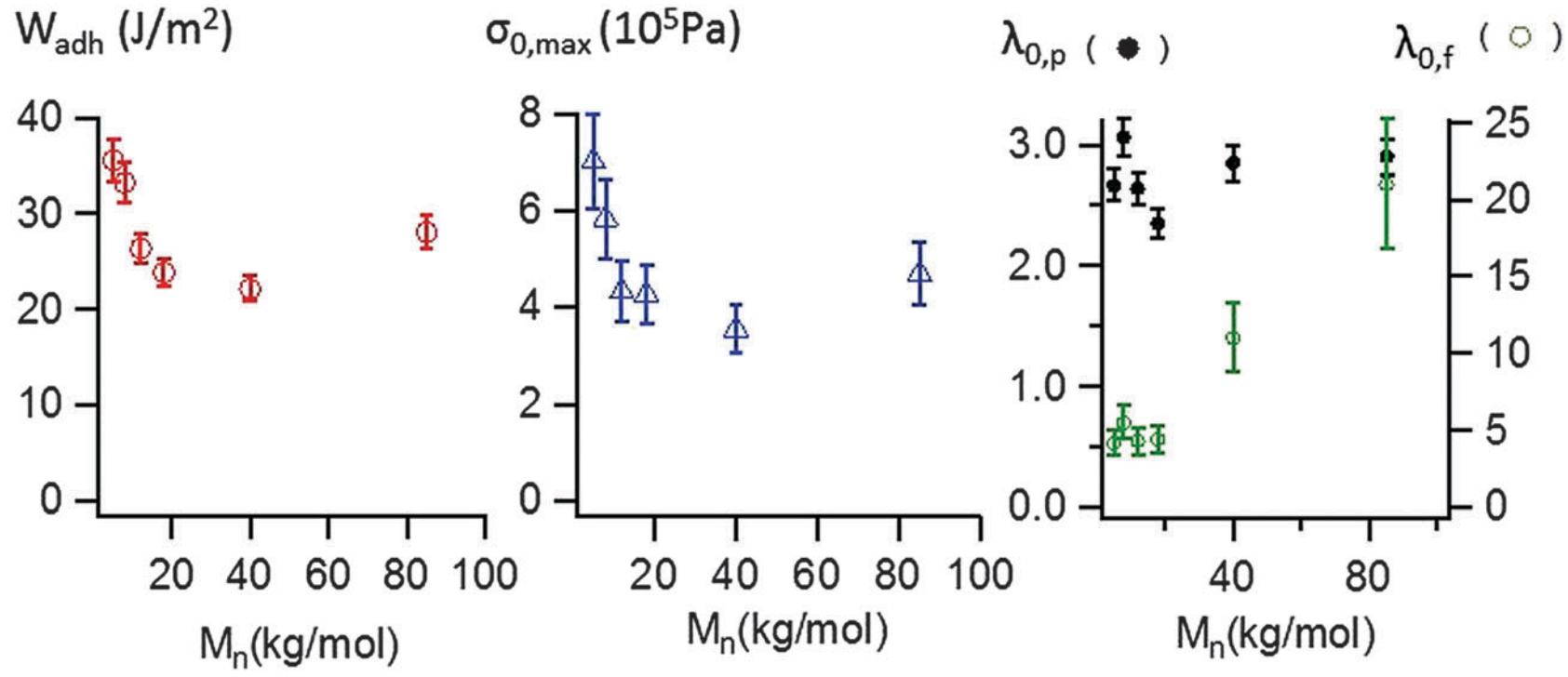

Fig. 8 Debonding energy ($W_{adh}$), maximal stress $\sigma_{0,max}$ and critical deformations ($\lambda_{0,p}$ and $\lambda_{0,f}$) of PnBA3U, measured using probe-tack tests at room temperature and at a debonding velocity $V_{deb}$ = 100 μm $s^{-1}$.

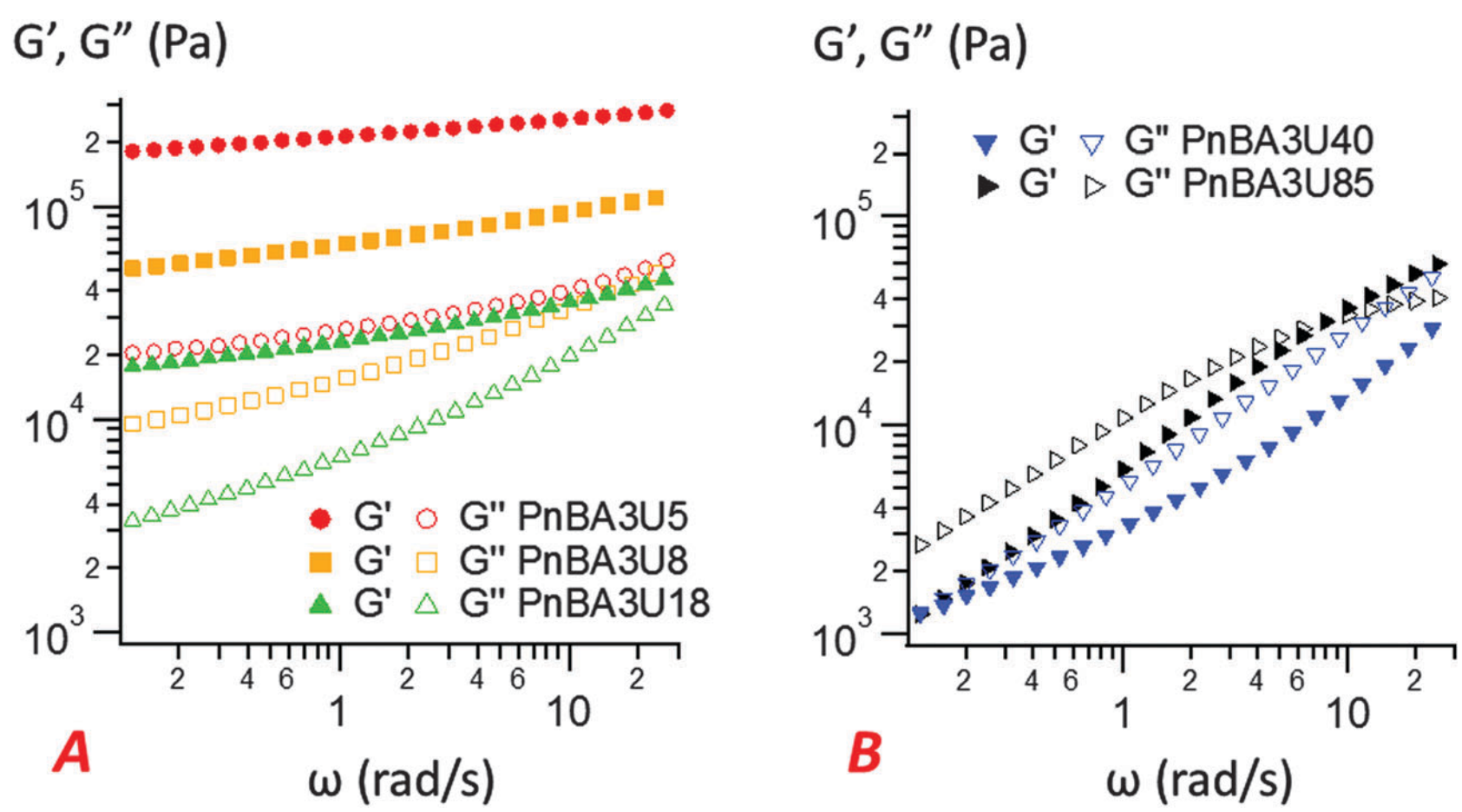

Fig. 9 Linear rheology curves at 25 °C for low (A) and high (B) molecular weight PnBA3U materials. The storage modulus $G'$ is represented by filled markers, and the loss modulus $G''$ is represented by unfilled markers (data from previous works[1]).

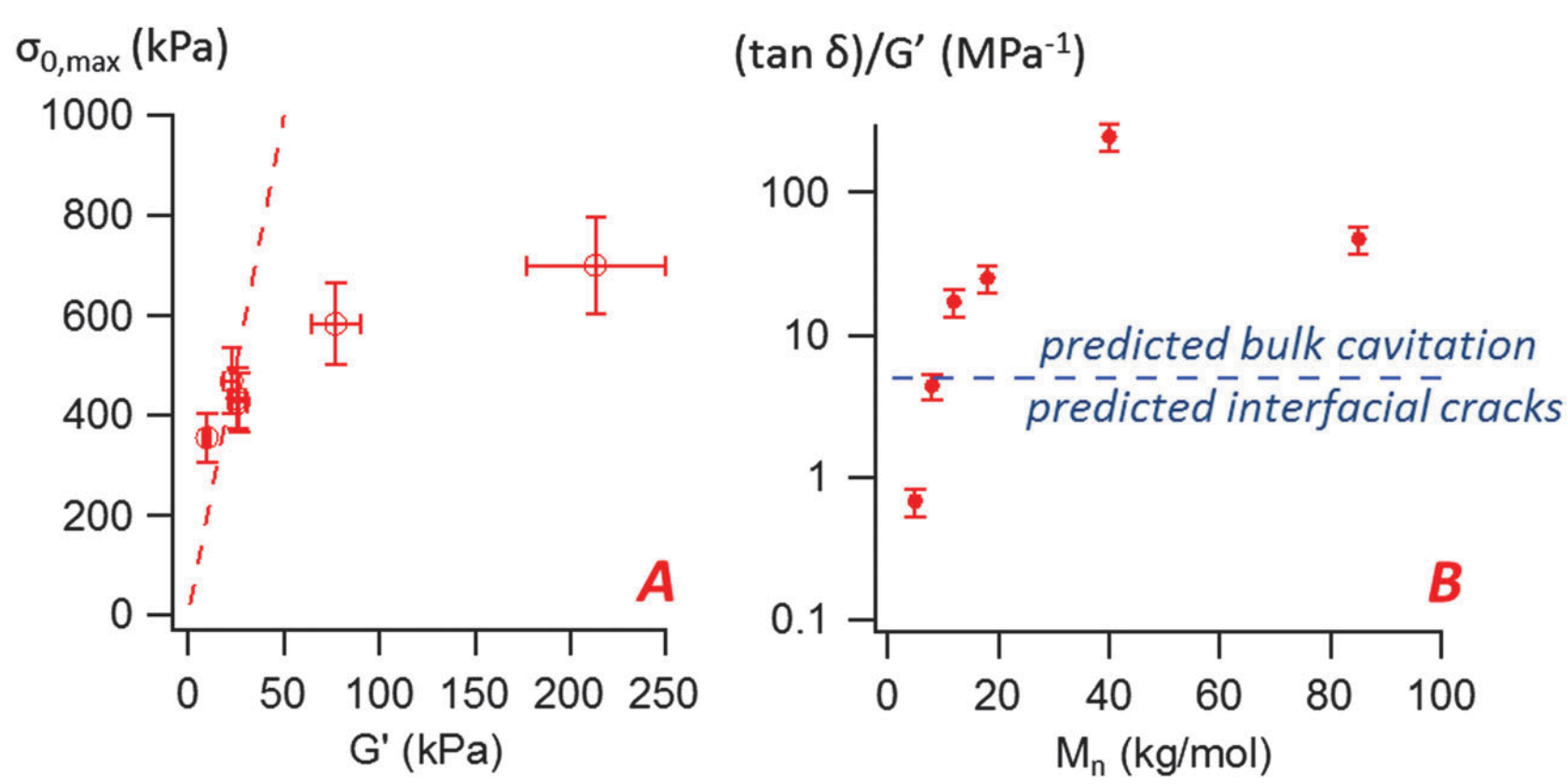

Fig. 10 (A) Variation of the maximal nominal stress $\sigma_{0,max}$ with the storage modulus $G'(\omega_c)$ measured at the representative angular frequency $\omega_c = 2\pi V_{deb}/h_0$ = 6.28 rad $s^{-1}$ (the dashed line represents $\sigma_{0,max} = 20G'(\omega_c)$), (B) variation of the ((tan $\delta$)/$G'$)($\omega_c$) ratio with the molecular weight for the PnBA3U materials. The straight line represents the constant $C$ proposed by Deplace et al.[32] to predict debonding by interfacial cracks or by bulk cavitation.

In addition to detaching at low stress peaks, the mode of growth of cavities observed for low molecular weight materials is also particularly unusual for elastic materials. The detachment of a solid substrate from a thin viscoelastic polymer layer usually occurs either by propagation of cracks at the interface between the substrate and the thin layer or by a mechanism of bulk cavitation, *i.e.* the growth of cavities into the adhesive layer (also observed for liquids). Inspired from theoretical works based on linear elastic fracture mechanics,[33] Deplace *et al.*[32] proposed that the transition between the two mechanisms could be predicted by the following ratio $C$:

$$C \approx k\mathcal{G}_0 \tan(\delta(\omega_c))/G'(\omega_c) \quad (12)$$

using the linear storage modulus $G'(\omega_c)$ and damping factor $\tan\delta(\omega_c)$ of the adhesive layer, measured at the average angular frequency characteristic of the adhesion test $\omega_c = 2\pi V_{deb}/h_0$, and an empirical prefactor $k$. $\mathcal{G}_0$ is here the resistance to crack propagation at vanishing debonding speed and depends mainly on the physical interactions at the interface.[34] For a fixed geometry (*i.e.* constant $k$) and similar surface chemistry (*i.e.* fixed $\mathcal{G}_0$), if $\tan(\delta(\omega_c))/G'(\omega_c) < C$, defects at the interface (such as air bubbles) are expected to propagate without significant deformation of the bulk. This mechanism of interfacial crack propagation generally leads to no macroscopic residue on the probe after the debonding step and the stress–strain curve is reduced to a high stress peak. If $\tan(\delta(\omega_c))/G'(\omega_c) > C$, the elastic energy stored in the adhesive layer is not sufficient to make interfacial cracks propagate and large strain in the bulk is expected before the detachment of the two layers. In Fig. 10, the ratio $\tan(\delta(\omega_c))/G'(\omega_c)$ of the PnBA3U materials is compared with the constant $C$ = 5 MPa$^{-1}$ measured by Deplace *et al.*[32] for conventional acrylic PSA on a steel surface. This prediction from linear viscoelasticity which works well for classical entangled adhesives[32,35,36] clearly fails for PnBA3U5: although $\tan(\delta(\omega_c))/G'(\omega_c) = 0.7 \pm 0.1$ MPa$^{-1}$ $\ll C$, the detachment of the steel probe occurred by bulk cavitation, as illustrated by the picture B of the locus of nucleation in Fig. 7. In other words linear viscoelasticity fails to predict the transition from cavitation to crack propagation. For $M_n > 8$ kg mol$^{-1}$, however, the observed mechanisms are consistent with the prediction based on the $(\tan\delta)/G'$ ratio.

For low molecular weight materials, the disagreement between our observations and the predictions suggests the existence of a pronounced softening mechanism which makes high local stress relax and prevents interfacial cracks from propagating. These large strain dissipative mechanisms are not captured by the linear dissipative response characterized by the damping factor $\tan(\delta)$ and the storage modulus $G'$. Therefore the non-linear rheology is required to better understand their unusual adhesive properties, as shown in the following part.

### Cavitation and extensional flow

In order to study insightfully the deformation of the supramolecular materials during adhesion tests, the debonding patterns were analyzed by the image analysis tools developed by Tanguy *et al.*[3] (see the Experimental part). For all PnBA3U materials, the cavitation process was characterized by measuring the total projected area $A_b$ covered by the cavities, the projected area $A_e$ of the walls between cavities (fibrils) as well as the sum of these two, the projected area $A_t$ of the cavitated disk shown schematically in Fig. 4. The ratios $A_b/A_0$, $A_e/A_0$ and $A_t/A_0$ are plotted *versus* the nominal stretch $\lambda_0$ in Fig. 11 and Fig. S3 (ESI†). For all $M_n$ values, $A_b/A_0$ increases first with $\lambda_0$ up to a maximum $(A_b/A_0)_{max}$ before decreasing sharply to 0 due to the rupture of the walls between cavities when the pressure equilibrates. In contrast the load bearing area, $A_e/A_0$ decreases continuously as the film is stretched. The reduction of the projected area of the walls is, however, not only due to the growth of the cavities but also explained by the reduction of the projected surface of the adhesive film, as revealed by the decrease of the ratio $A_t/A_0$ (for $\lambda_0 < 2$).

While the decrease of the total area $A_t$ of the film is qualitatively similar for all materials, the growth rate of the bubbles is different from one material to another. Although some scatter is observed for high molecular weight materials, the projected area of the cavities when the pressure equilibrates varies qualitatively with the viscoelastic moduli of the thin films (see Fig. 9 and 11). When $M_n$ increases from 5 to 40 kg mol$^{-1}$ $(A_b/A_0)_{max}$ decreases from 55 to 30% and then increases again to 40% for $M_n$ = 85 kg mol$^{-1}$. The presence of cavities in the adhesive layer significantly reduces the load-bearing area $A_e$,

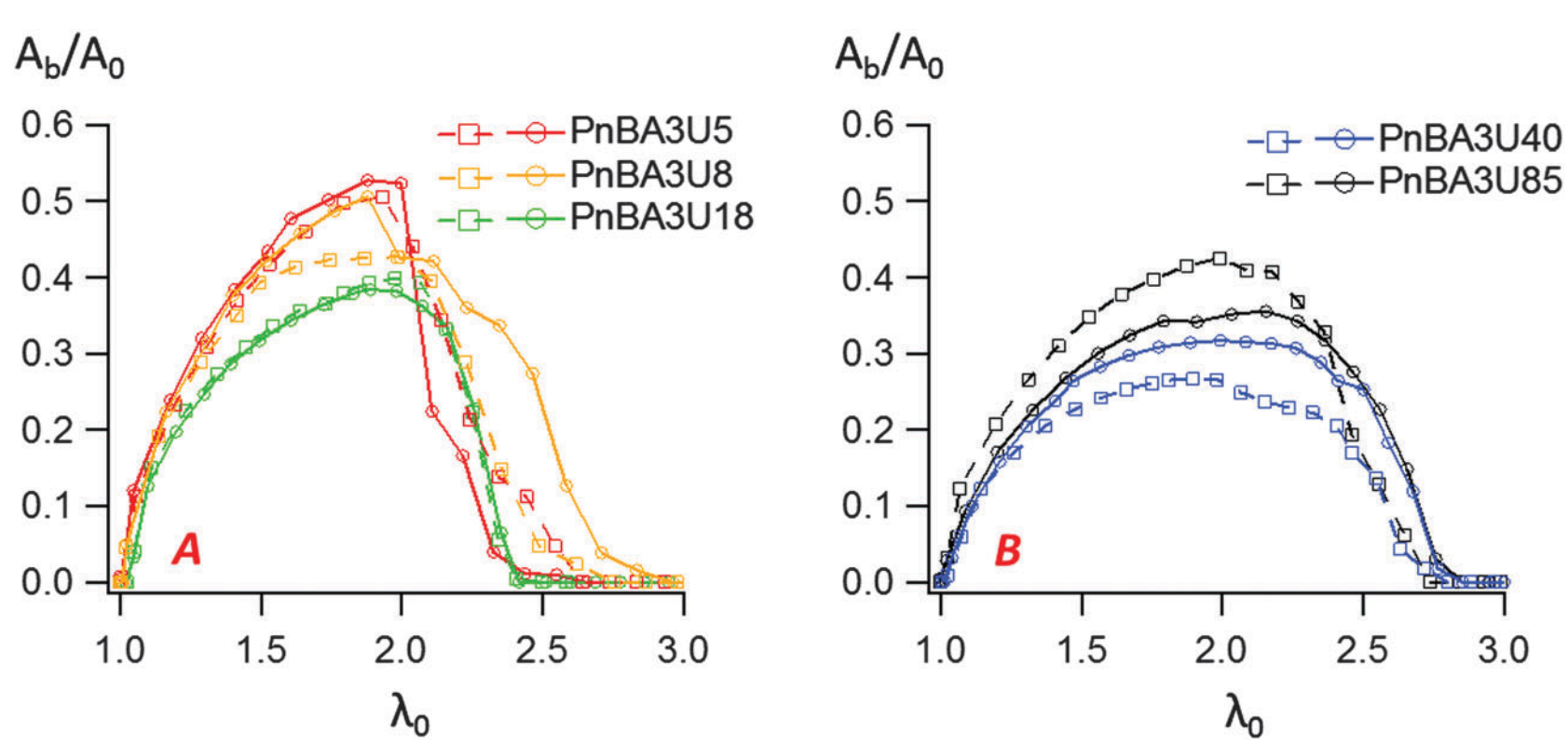

Fig. 11 Variation of the ratio $A_b/A_0$ with the nominal strain for low (A) and high (B) molecular weight PnBA3U materials. Two experimental curves are shown for each material to indicate the reproducibility of the results.

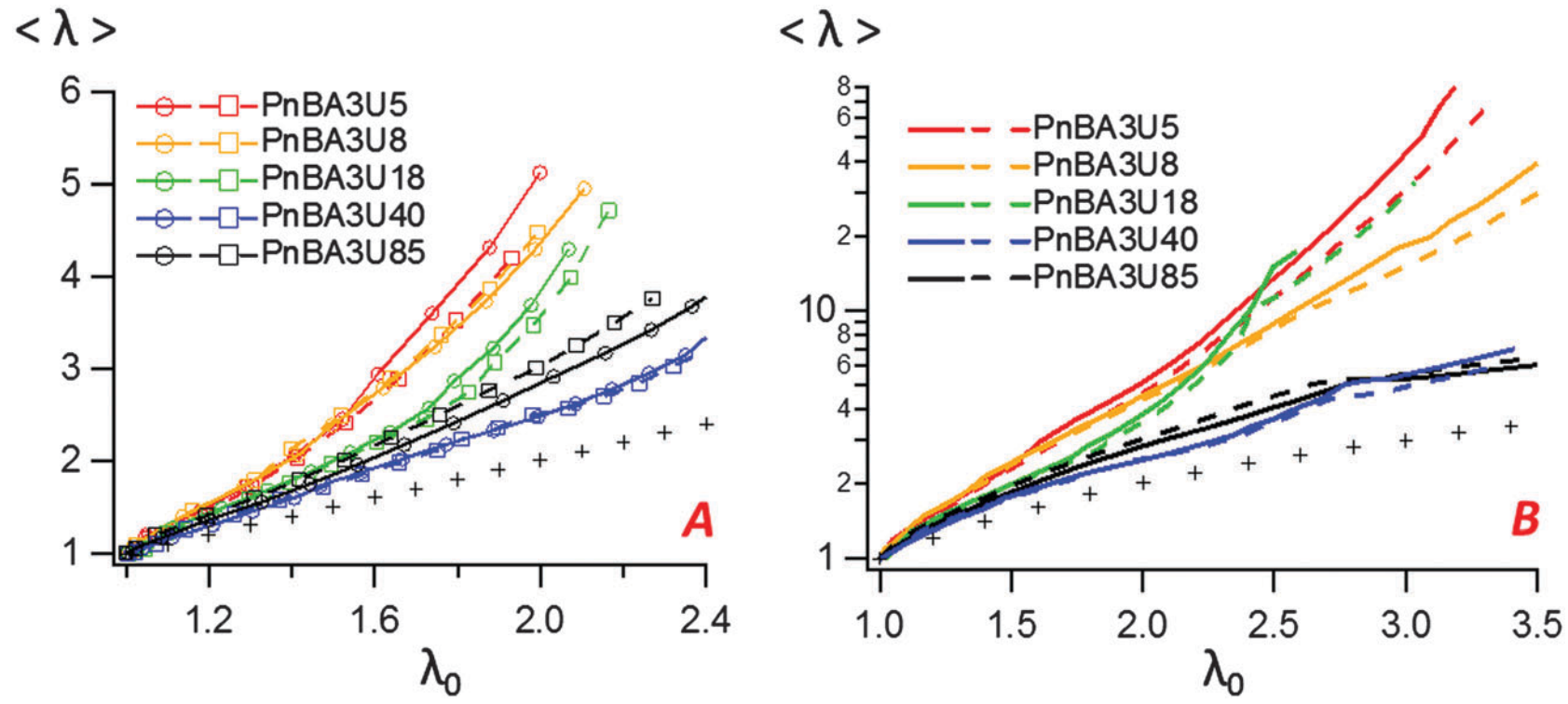

Fig. 12 Variation of the local averaged deformation $\langle\lambda\rangle$ with the nominal strain for different PnBA3U materials before (A) and after (B) the equilibration of the pressure. Two experimental curves are shown for each material. On the left, the plot focuses on the variation of the local elongation during the phase of the growth of bubbles (before the coalescence of bubbles). The line that is drawn with crosses represents the hypothetical case where $\langle\lambda\rangle = \lambda_0$.

*i.e.* the walls between cavities that are stretched during the debonding step. The consequence of the growth of cavities on the deformation of the fibrils can be captured by plotting the local elongation of the walls $\langle\lambda\rangle$ *versus* the nominal elongation $\lambda_0$ (see Fig. 12A). For all molecular weights, $\langle\lambda\rangle$ is initially equal to $\lambda_0$ and then increases non-linearly with $\lambda_0$. At a fixed nominal stretch $\lambda_0 > 1$, $\langle\lambda\rangle$ varies significantly from one PnBA3U to another and as expected, $\langle\lambda\rangle$ is maximal for low molecular weight materials and is minimal for PnBA3U40. As shown in Fig. 12B, the thinning mechanisms of the fibrils can also be characterized after the equilibration of the pressure. Although the experimental data are more scattered, this mechanism is clearly more pronounced for the supramolecular gels compared to that for the viscoelastic liquids: at $\lambda_0 = 3.5$, $\langle\lambda\rangle$ is lower than 10 for $M_n \geq 40$ kg mol$^{-1}$ while $\langle\lambda\rangle$ is higher than 30 for $M_n \leq 18$ kg mol$^{-1}$.

In order to characterize the behavior of the PnBA3U materials at large deformations, the true effective stress $\sigma_e$ was calculated from the projected area of fibrils and the force $F_T$, as previously explained. The variation of $\sigma_e$ with the effective strain $\langle\lambda\rangle$ before the equilibration of pressure is shown in Fig. 13. All $\sigma_e = f(\langle\lambda\rangle)$ curves show a similar shape with a stress peak at low elongation and then a true stress plateau up to $\langle\lambda\rangle \sim 4$. The level of the stress plateau and the stress peak decrease in the molecular weight range from 5 to 40 kg mol$^{-1}$ and slightly increase for $M_n$ = 85 kg mol$^{-1}$. In classic tensile experiments, the stress plateau at large deformation is usually observed for non-cross-linked and lightly entangled polymers and is characteristic of a liquid flow under stress. In Fig. 13, the stress plateau suggests the motion of the supramolecular polymer chains within the walls between cavities and thus, the low molecular weight PnBA3U materials soften when the strain amplitude increases as a yield stress fluid would.[37]

The variation of the Hencky strain rate $\dot{\varepsilon}_H$ with the nominal strain $\lambda_0$ before and after the equilibration of pressure is illustrated with PnBA3U5 and PnBA3U85 in Fig. 14A. At the beginning of the adhesion step, the nucleation step induces a sharp increase of $\dot{\varepsilon}_H$ for both materials and then, as the thin film is stretched at a fixed nominal deformation rate, $\dot{\varepsilon}_H$ remains roughly constant for PnBA3U5 while $\dot{\varepsilon}_H$ slightly decreases for PnBA3U85. After the equilibration of pressure, for the remaining fibrils between the steel probe and the glass slide, $\dot{\varepsilon}_H$ still decreases for $M_n$ = 85 kg mol$^{-1}$ while $\dot{\varepsilon}_H$ increases for $M_n$ = 5 kg mol$^{-1}$. It is worthwhile noting that, at a fixed $\lambda_0$, $\dot{\varepsilon}_H$ is much higher for the supramolecular gels than for the viscoelastic liquids, particularly after the coalescence of cavities. Although the deformation of the thin film occurred at a fixed probe pullout velocity, this difference of the Hencky strain rate between low and high molecular weight materials reveals that fibrils created in the low molecular weight materials localize the strain and are much less stable under stress compared to that for the high Mw ones. In the case of the supramolecular gels, the weak cohesion of the fibrils, which would account for the sharp decrease of $\sigma_0$ observed after the equilibration of pressure (see Fig. 5), would be a direct consequence of the sharp strain softening of the supramolecular gels at large deformations. As the cavities nucleate, the supramolecular gels soften at the edges of the cavities, where local stresses are higher. This local softening mechanism accelerates the propagation of cavities until the rupture of the walls.

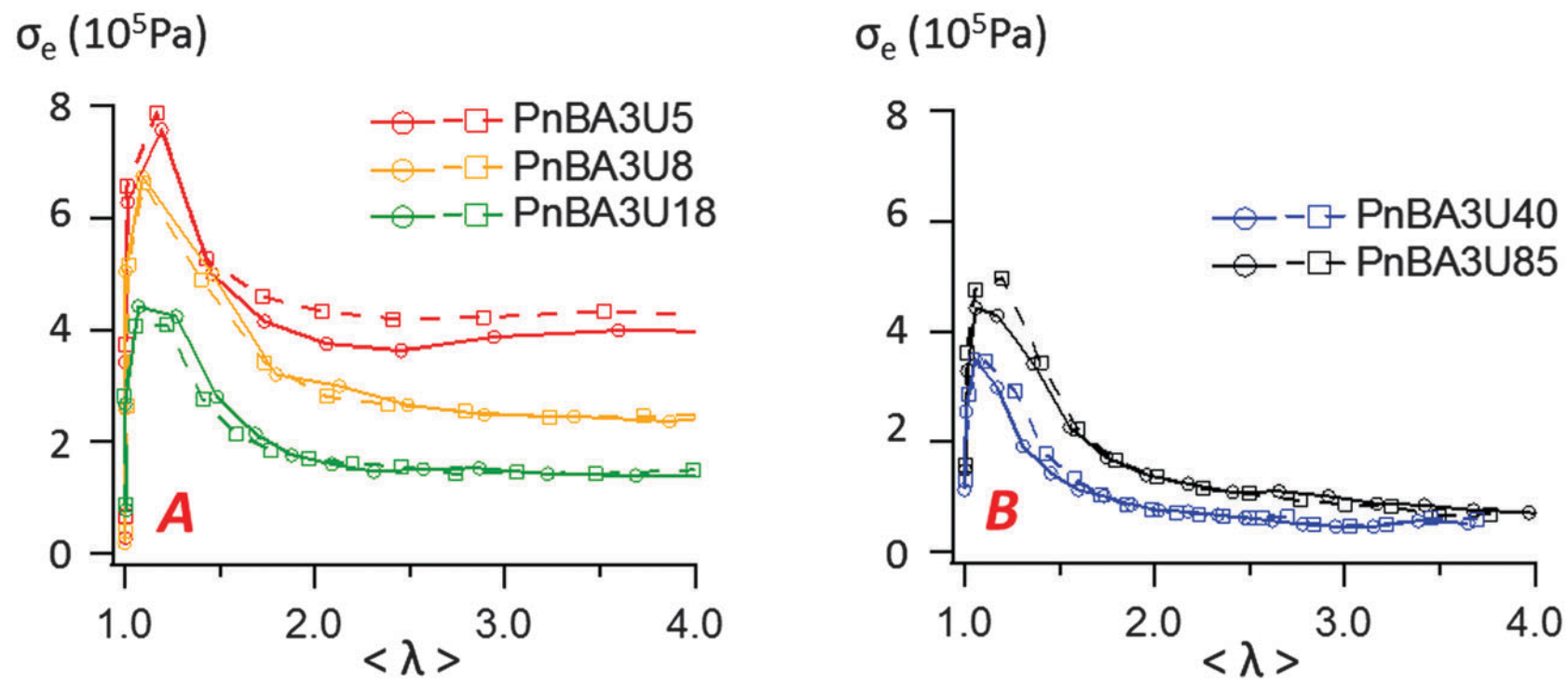

Fig. 13 Variation of the effective stress with the effective strain $\langle\lambda\rangle$ when bubbles grow in the adhesive layer for low (A) and high (B) molecular weight PnBA3U materials. Two experimental curves are shown for each material.

At large strain, the absence of strain-hardening usually observed for soft acrylic adhesives[38,39] is confirmed by the variation of the local extensional viscosity $\eta^+$ during the probe-tack test. The variation of $\eta^+$ with $\lambda_0$ is illustrated by PnBA3U5 and PnBA3U85 in Fig. 14B. Before the equilibration of pressure, $\eta^+$ decreases for both materials while the walls between cavities are stretched. After the coalescence of bubbles, the extensional viscosity slightly increases for high molecular weight PnBA3U materials while some scatter is observed for low $M_n$ PnBA3U materials. For PnBA3U85, the increase of $\eta^+$ is consistent with the reduction of $\dot{\varepsilon}_H$ (see Fig. 14A) as usually observed for non-Newtonian liquids. For low molecular weight PnBA3U materials, the drop of $\eta^+$ highlights their liquid-like behavior at large deformation which accounts for the strain localization and, thus, the cohesive failure observed for these materials.

In order to shed light on the origin of the non-linear rheology of both materials, the local extensional viscosity $\eta^+$ measured during the probe tack test can be compared with the limiting linear viscoelastic (LVE) extensional viscosity $\eta^+_{LVE}$. In the case where $\dot{\varepsilon}_H$ is constant, $\eta^+_{LVE}$ can be calculated from the relaxation modulus $G(t)$:

$$\eta^+_{LVE}(t) = 3\int_0^t G(x)\mathrm{d}x \tag{13}$$

For both materials, $G(t)$ was estimated from a multimode Maxwell fit to the LVE data shown in Fig. 9 (see Fig. S4 for Maxwell parameters, ESI†) and, although $\dot{\varepsilon}_H$ is roughly constant during probe-tack tests, $\eta^+_{LVE}$ was calculated using eqn (13). As observed in Fig. 14B, $\eta^+$ does not follow the LVE envelope: in the fibrillation zone, $\eta^+$ is lower than $\eta^+_{LVE}$ for PnBA3U5 while $\eta^+$ is higher than $\eta^+_{LVE}$ for all values of $\lambda_0$ for PnBA3U85.

For low molecular weight materials, the high LVE extensional viscosities reflect the elastic plateau observed at low frequency in the LVE data. This plateau was previously shown to result from the hexagonal structure of supramolecular filaments into bundles.[1,2] These filaments are themselves formed by the self-assembly of tri-urea cores and surrounded by PnBA side chains. Since each PnBA3U linear chain is monofunctional, interactions between two neighboring filaments (which can be seen as large colloidal objects) are limited to van der Waals interactions between unentangled poly(butylacrylate) brushes. The discrepancy between $\eta^+$ and $\eta^+_{LVE}$ suggests that the breakup of this supramolecular nanostructure, induced by high local stresses, is responsible for the flow of the material at large deformation. Further studies are needed to determine the precise molecular mechanisms behind the softening mechanism. The scission of the supramolecular self-assembly is probably favored by the polarity of the PnBA monomers (relative to polyisobutylene, for example[17]) which are able to interact with urea groups reducing the strength of interactions between tri-urea cores. On the other hand, recent studies on "supersoft" and "superelastic" networks in the melt state[40] suggest that the bottlebrush shape of these aggregates with unentangled arms provides them high mobility and thus, the observed strong softening of the material at high strain may not require the scission of the tri-urea sticker self-assembly. In-depth investigation of the length and the stiffness of these aggregates as well as monitoring by SAXS of the microstructure under stress would be required to gain more insight on the contribution of the supramolecular dynamics to the behavior of PnBA3U at large deformation.

For high molecular weight materials, *i.e.* rather dilute stickers, the opposite trend is interestingly observed ($\eta^+ > \eta^+_{LVE}$ for all $\lambda_0$). The Hencky strain rates measured during probe-tack tests ($\dot{\varepsilon}_H \leq 2\ \mathrm{s}^{-1}$) are lower than the inverse disentanglement time ($1/\tau_d \sim 8\ \mathrm{s}^{-1}$ for PnBA3U85). By analogy with linear homopolymers, $\eta^+$ and $\eta^+_{\mathrm{LVE}}$ should be similar for PnBA3U85. The difference between these expectations and the experimental results reveal the impact of small clusters of stickers on the motion of the polymer chains at large deformation while the effect of stickers is hardly discernible at small strain. A qualitatively opposite discrepancy was recently observed in a recent study by Shabbir and coworkers on entangled poly(butylacrylate–acrylic acid) melts.[14] In extensional rheology, their work revealed an upward deviation of $\eta^+$ from the LVE envelope at low density of stickers per chain (~3%) and at low strain rates (below the inverse of the Rouse time, which is the threshold strain rate where strain hardening is typically observed for linear entangled polymers). In linear rheology, the effect of the stickers was limited to the terminal zone where, as observed in our own study,[1] slopes of 1 and 2 were not reached for the elastic and loss moduli, respectively. However, these authors detected a strain hardening mechanism, *i.e.* a strong increase of the extensional viscosity with time and they attributed it to "trapped" polymer segments, *i.e.* poly(butylacrylate) segments linked to different poly(acrylic acid) nanodomains.[41] The monofunctionalization of PnBA3U chains prevents the existence of such trapped segments between the supramolecular rods which act more as fillers rather than as physical cross-linkers. The parallel between both systems underlines the necessity for coupling entanglements with supramolecular clusters to get strain hardening. Incorporation of a chain extender would be required to trigger the strain hardening of the PnBA3U material stabilizing the fibril extension and favoring an adhesive mode of fracture.

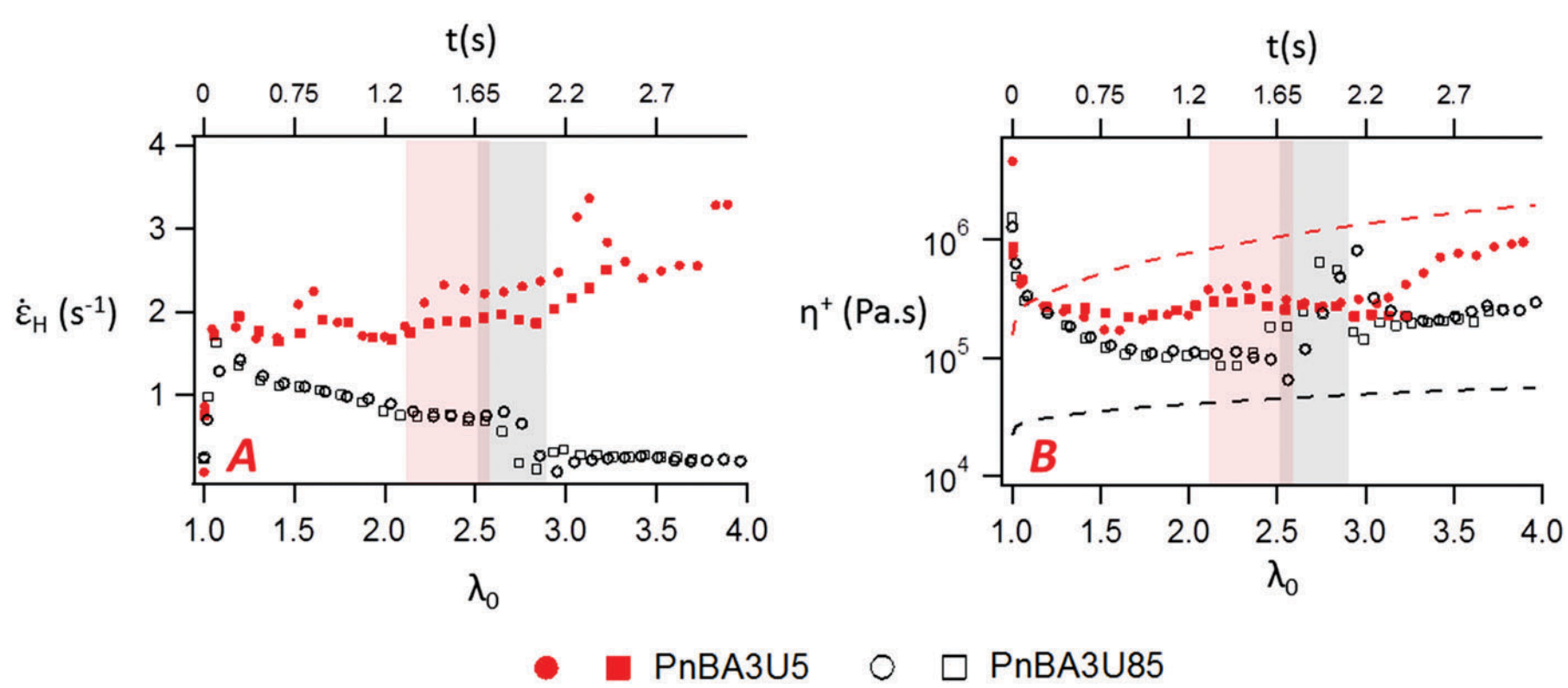

Fig. 14 Variation of the Hencky strain rate (A) and the local extensional viscosity (B) with the nominal strain and time for PnBA3U5 (red markers) and PnBA3U85 (black markers). The highlighted regions corresponds to the phase of the equilibration of pressure, where $\dot{\varepsilon}_H$ and $\eta^+$ are scattered and thus, difficult to interpret. Two experimental curves are shown for each material. In Fig. 13B, the red and black dashed lines are the limiting linear viscoelastic extensional viscosities of PnBA3U5 and PnBA3U85, respectively.

In summary, two processes must be taken into account to understand the influence of the molecular weight on the mechanisms of debonding of PnBA3U. As expected from the linear rheology, the decrease of $M_n$ strengthens the cohesion of the adhesive film, as suggested by the increase of the stress plateau with the density of stickers in Fig. 13. This process is counter-balanced by a mechanism of localization of strains induced by the softening character of the PnBA3Us. The softening of the bulk at the edges of the cavities provokes high local strain rates (but low stresses) which make their growth in the bulk easier until the walls break and the probe detaches leaving residues. A clear manifestation of this strain localization effect is also seen in Fig. 5 for values of $\lambda_0 > 3$, *i.e.* after the pressure has equilibrated, the force decreases much slower for the high molecular weight materials than that for the low molecular weight materials indicating that the strain localizes much less for the entangled melts compared to that for the supramolecular non-entangled polymers demonstrating the crucial effect of strain hardening in debonding processes.

The competition between these effects explains why the adhesive properties are relatively similar for all materials and poor compared to classical entangled and slightly chemically crosslinked soft adhesives.[42] While the highest debonding energy measured in our study is around 37 J m$^{-2}$, the debonding energy of commercial PSAs are usually measured between 100 and 1000 J m$^{-2}$ for such a layer thickness, debonding rate and temperature.[38,43] These performances mainly come from the strain hardening capability of the material which avoids the strain localization in the fibrils and allows large deformations of the adhesive layer before debonding occurs. For similar experimental conditions, the critical strain $\lambda_{0,\mathrm{p}}$ of acrylic adhesive varies typically between 4 and 10 in the literature while $\lambda_{0,\mathrm{p}}$ remains lower than 3.3 for the PnBA3U systems.

## Conclusion

In this study, the adhesion properties of tri-urea center-functionalized poly(butylacrylate) were investigated by probe-tack tests at room temperature in the molecular weight ($M_n$) range between 5 and 85 kg mol$^{-1}$. While an elastic response was previously reported in the linear regime for low $M_n < 40$ kg mol$^{-1}$, a cohesive mode of fracture, which is a feature of liquids, was observed for all molecular weights. In order to highlight the role of non-linear rheology in these adhesion tests, the growth of cavities was carefully analyzed using an Image analysis tool and a local Hencky strain rate could be calculated from the projected area of fibrils. As the force sensor was synchronized with the camera, the effective stress in the walls of the cavities was also calculated and finally a local extensional viscosity could be determined.

The sharp decrease of the local extensional viscosity for all materials implies a softening mechanism of the supramolecular material at large deformation which is classical for entangled polymer melts but resembles that of yield stress fluids for the low $M_n$ viscoelastic gels. For the lower molecular weight materials, the strain-softening character of these materials also induced a mechanism of localization of stresses akin to fracture which accelerates the propagation of cavities until the detachment of the probe. All these factors explain the low adhesion energy against steel ($W_{\mathrm{adh}} < 40$ J m$^{-2}$) measured for all $M_n$ values.

The present work raises important points for the molecular design of innovating supramolecular adhesives from supramolecular center-functionalized polymers. Previous studies reported that their viscoelastic properties were highly tunable and made them interesting for applications where highly dissipative properties are required over a precise time range.[17,44] However, their softening character is not compatible with Pressure Sensitive Adhesives which ideally require clean removal from the substrate surface. Modification of the chemical architecture of the polymer chains would be required to prevent the polymer chains from flowing at large deformation.